\documentstyle[12pt,epsf,graphics]{article}
\def\be{\begin{equation}}
\def\ee{\end{equation}}
\def\bear{\begin{eqnarray}}
\def\eear{\end{eqnarray}}
\begin{document}
\title{Instability of the Time-Dependent Ho\u rava-Witten Model} 
\author{ B. Cuadros-Melgar$^1$, C.E. Pellicer$^2$ \\ \\
$^1$Departamento de F\'isica, Universidad de Santiago de Chile,\\
Casilla 307, Santiago, Chile\\
$^2$Instituto de F\'\i sica, Universidade de S\~ao Paulo, \\
C.P.66.318, CEP 05315-970, S\~ao Paulo, Brazil}
\date{}
\maketitle

\begin{abstract}
We consider scalar perturbations in the time-dependent Ho\u rava-Witten
Model in order to probe its stability. We show that during the
non-singular epoque the model evolves without instabilities until it
encounters the curvature singularity where a big crunch is supposed to
occur. We compute the frequencies of the scalar field oscillation during
the stable period and show how the oscillations can be used to prove the
presence of such a singularity.
\end{abstract}

\begin{center}
\end{center}

\newpage

\section{Introduction} 
The classical Big-Bang model based on Einstein's General Relativity does
not explain the origin of the initial singularity, which is expected to be
understood at a higher dimensional string theory model.

It is known that the Ho\u rava-Witten theory \cite{hw} relates
11-dimensional supergravity on the orbifold $S^1/Z_2$ with strongly
coupled heterotic $E_8 \times E_8$ string theory. Then, this link suggests that
as one probes higher energies, our 4-dimensional universe would first go
through an intermediate regime where the orbifold dimension becomes
visible, thus making our world appear 5-dimensional \cite{lukas,banks}, and only
at energies of order the string scale the universe would look
11-dimensional. 

The intermediate 5-dimensional energy regime has led to the conjecture that
our universe is a brane in a higher dimensional world. These Ho\u
rava-Witten inspired ``braneworlds'' have a topology defined by a line
times a non-compact space with two branes at the boundaries of the line
element, which is usually characterized as the orbifold $S^1/Z_2$. Due to the compactness of the orbifold direction, one brane has positive tension while the other has the opposite negative one.  By
construction these spacetimes are supersymmetric and this unbroken
supersymmetry of the static background solution guarantees the local
stability of the model \cite{stab}.

However, the situation changes when the background solution becomes time-dependent. The idea of having moving branes has naturally led to the
suggestion that the hot big bang is the result of a brane collision
\cite{ekp}, or even that our universe undergoes an endless sequence of
cosmic epochs which begins with a universe expanding from a big bang and
ending with a contraction to a big crunch \cite{cycle}. 

Colliding braneworld scenarios offer an alternative point of view for many
long-standing problems and puzzles such as those connected with the
existence of gravitational shortcuts (apparently faster than light signals) \cite{abdetal}. 
These issues refer to the different perspective of higher dimensional theories and conventional 4-dimensional ones. In fact, the proposed observational tests of braneworld scenarios must be able to distinguish between higher and lower dimensional effects. Moreover, the former effects must appear as corrections of the latter in these models. 

One of the approaches used to study these collisions is through low
energy supergravity limits. Since we lack a complete M-theory formulation, the solutions come from the 5-dimensional supergravity theory obtained after a dimensional reduction of the higher dimensional scenario. In the case of Type II-B theory exact solutions to this problem have been
found in \cite{bwcoll1}, where the motion of the D3 branes towards one
another leads to the complete disappearance of the universe in a spacetime
singularity.

In the context of heterotic brane models an exact solution of the equations of motion derived from the same Ho\u rava-Witten lagrangian has been found \cite{chen}. As in the previous case, 
the setup obtained from a dimensional reduction to 5 dimensions evolves to a singularity where the entire spacetime is
annihilated. This can be taken as an indication of an inherent classical
instability in braneworld models consisting of a positive and a negative tension
brane. However, it has recently been proposed a way to avoid the annihilation by introducing an arbitrarily small amount of matter on the negative tension brane \cite{jeanluc}. This alternative scenario is specially appealing for the ekpyrotic and cyclic universe models \cite{ekp,cycle} due to the momentary disappearance of the orbifold dimension caused by the collision and the subsequent bounce after which both branes continue to expand into the future.

In this paper we study the stability of the solution in \cite{chen} by
introducing a scalar perturbation. General perturbations of solutions to
the Einstein equations are very well known in the literature, see
e.g. \cite{x} and further references therein. In the present case we deal
with perturbations of time-dependent solutions, a more delicate
issue \cite{shao}.  Although we know from
previous discussions that there exists an instability \cite{bwcoll1,chen}, our motivation is to check the
possibility of obtaining some information about the signatures we would
expect before the spacetime collapses. 

The paper is organized as follows. In section 2 we describe the time-dependent Ho\u rava-Witten solution under study. In the next section we
introduce the scalar field perturbation, we solve the corresponding
Klein-Gordon equation in the bulk, and we discuss the field evolution in
the setup. Finally, we present our conclusions in section 4.

\section{The Time-Dependent Ho\u rava-Witten Model}

One of the most popular models connecting M-theory with phenomenology is the
Ho\u rava-Witten Model \cite{hw}. In this setup the 11-dimensional
spacetime is the product of a Calabi-Yau compact manifold and a
5-dimensional spacetime consisting of two parallel 3-branes or domain
walls, one with negative tension and the other with positive one. When
we perform a dimensional reduction from eleven to five dimensions, we
obtain a 5-dimensional supergravity theory which admits an exact static
supersymmetric solution of the form \cite{lukas} 
\begin{equation}\label{static}
ds_5 ^2 = \tilde H (-dt^2 + d \vec x ^2) + \tilde H^4 d\tilde y ^2 \, ,
\end{equation}
where
\begin{equation}\label{st-param}
\tilde H = 1+ \tilde k |\tilde y|\, , \quad \phi = -3 \log \tilde H \, ,
\end{equation}
and $\tilde k^2 = 2m^2/3$ is a constant, while $\phi$ is a scalar field
characterizing the size of the internal Calabi-Yau space. The equations of
motion for the metric and $\phi$ can be obtained from the Lagrangian 
\begin{equation}\label{lag}
{\cal L}_5 = \sqrt{-g}\, (R-\frac{1}{2} (\partial \phi)^2 -m^2
e^{2\phi})\, . 
\end{equation}
In this scenario a second domain wall is introduced at $y=L$, taking $y$ as
a periodical coordinate with period $2L$ and making the $Z_2$
identification, $y \leftrightarrow -y$. 

The model based upon this solution is not physically realistic because it is
static, while our universe is time-dependent. Led by this motivation many
attempts to introduce dynamics in this model have been made. Remarkable
examples are the Ekpyrotic Model \cite{ekp}, where the big bang is regarded 
as the result of the collision of an external brane with our universe, and
the Cyclic Universe \cite{cycle}, where the distance between the branes
oscillates in time.  

However, recently Chen {\it et al.} \cite{chen} found that the lagrangian
(\ref{lag}) also admits a time-dependent 3-brane solution given by 
\begin{equation}\label{metric}
ds_5 ^2 = H^{1/2} (-dt^2 + d\vec x ^2) + H dy^2 \, ,
\end{equation}
with 
\begin{equation}\label{param}
H= ht + k |y| \, , \quad \phi = - \frac{3}{2} \log H \, ,
\end{equation}
where $k^2 = 8m^2/3$ and $h$ is an arbitrary constant.

If we turn off the time-dependence in this solution, {\it i.e.}, set $t=0$, we can go back to the time-independent solution (\ref{static})-(\ref{st-param}) by performing the coordinate transformation $y=\frac{k}{4}\tilde y^2$ and rescaling the constant $c_0$ of the original work of Lukas {\it et al.} \cite{lukas}, where $\tilde H=\tilde k |\tilde y| + c_0$. In fact, this transition between the solutions can be better visualized if we set $c_0=0$, which is always possible since it is an integration constant. 

The solution (\ref{metric})-(\ref{param}) represents a bulk spacetime with two 3-branes with negative and positive tensions as boundaries
at $y=0$ and $y=L$, respectively. The interval $0 \leq y \leq L$ is of $S^1/Z_2$ type, where $S^1$ occupies the strip $-L\leq y\leq L$ and the $Z_2$ symmetry corresponds to the identification $y\leftrightarrow y$. By choosing $h<0$ as in \cite{bwcoll2} we notice that $H$ is a positive but decreasing quantity when $t<0$. The spacetime is bounded in the range $0\leq y \leq L$, but the proper length of this interval is time-dependent. Thus, the separation between the branes decreases as $t$ becomes positive, {\it i.e.}, the universe is contracting. When $t$ turns out to be positive, $H$ vanishes along $-ht=ky$. This line represents a singularity which first develops on the negative tension brane ($y=0$) at $t=0$ and travels along the orbifold dimension reaching the positive tension brane ($y=L$) after a finite time $t=kL/(-h)$. The spacetime can not be extended beyond
this point. Thus, before the branes effectively collide the power-law singularity wraps the whole spacetime.  

Although we know that an instability exists in the present model, we still want to check the possibility of describing its existence before the spacetime
collapses. 

\section{The Scalar Perturbation}

We consider a perturbation in the form of a scalar field obeying the
massive Klein-Gordon equation 
\begin{equation}\label{kg0}
\Box \Phi = m^2 \Phi \, ,
\end{equation}
or equivalently
\begin{equation}\label{kg}
\frac{1}{\sqrt{-g}} \partial_M (\sqrt{-g} \, g^{MN} \partial_N) \Phi =
m^2 \Phi \, . 
\end{equation}

Using the metric (\ref{metric}) in spherical coordinates and the
derivatives of $H$, $\partial_t H = h$ and $\partial_y H = k \,
\hbox{sgn}(y)$, Eq.(\ref{kg}) can be rewritten as
\bear\label{kg-1}
\left\{ - H^{-1/2} \partial_t ^2 - h H^{-3/2} \partial_t + H^{-1/2}
\partial_r ^2 + \frac{2}{r} H^{-1/2} \partial_r + \frac{H^{-1/2}}{r^2}
\times \right. \qquad \qquad \qquad \nonumber \\
\left. \times \left[ \frac{1}{\sin \theta} \partial_{\theta}
  (\sin \theta \partial_{\theta}) + \frac{1}{\sin ^2
    \theta}\partial_{\phi} ^2 \right] + H^{-1} \partial_y ^2 +
\frac{k}{2} H^{-2} \hbox{sgn}(y) \partial_y - m^2 \right\}\Phi = 0 \, . 
\eear
Now we decompose the scalar field $\Phi$ as
\be
\Phi(t,r,\theta,\phi,y) = Z (t,r,y) Y_{\ell m}(\theta,\phi) \, ,
\ee
where the spherical harmonics obey the equation
\be
\frac{1}{\sin \theta} \partial_{\theta} (\sin \theta \partial_{\theta}
Y_{\ell m}) + \frac{1}{\sin ^2 \theta} \partial_\phi ^2 Y_{\ell m} =
-\ell (\ell +1) Y_{\ell m} \, .
\ee
Thus, the function $Z (t,r,y)$ satisfies the equation
\bear\label{kg-2}
-\partial_t ^2 Z - \frac{h}{H}\partial_t Z + \partial_r ^2 Z +
\frac{2}{r} \partial_r Z - \frac{\ell (\ell +1)}{r^2} Z + H^{-1/2}
\partial_y ^2 Z + \nonumber \\
+\frac{k}{2} H^{-3/2} \hbox{sgn}(y) \partial_y Z -
m^2 H^{1/2} Z =0 \, .
\eear

\begin{figure*}[htb!]
\begin{center}
\leavevmode
\begin{eqnarray}
\epsfxsize= 8.0truecm\rotatebox{0}
{\epsfbox{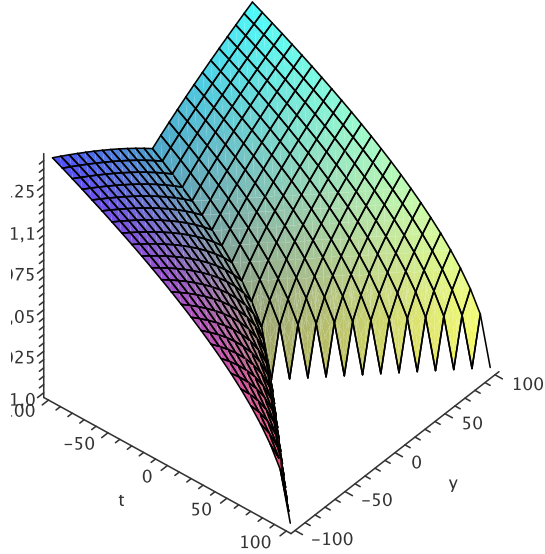}}\nonumber
\end{eqnarray}
\caption{Potential $\alpha^2 + m^2\, H^{1/2}$ for $\alpha^2=1$, $m=0.1$.}
\label{pot}
\end{center}
\end{figure*}

In order to separate the radial part we write $Z$ as $Z(t,r,y) = \Psi
(t,y) R(r)$. Substituting back into (\ref{kg-2}) we obtain 
\be\label{radial}
\partial_r ^2 R + \frac{2}{r} \partial_r R + \left( \alpha^2 -
\frac{\ell (\ell +1)}{r^2} \right) R =0 \, ,
\ee
and
\be\label{Psi}
\partial_t ^2 \Psi + \frac{h}{H} \partial_t \Psi - \frac{1}{\sqrt{H}}
\partial_y ^2 \Psi - \frac{k\, \hbox{sgn}(y)}{2 H^{3/2}} \partial_y \Psi
+ (\alpha^2 +m^2 H^{1/2}) \Psi = 0 \, ,
\ee
where $\alpha^2$ is a constant.

The solution to equation (\ref{radial}) is given by
\be
R(r) = \frac{A}{\sqrt{r}} \, J\left( \frac{1}{2} + \ell , \alpha r
\right) + \frac{B}{\sqrt{r}} \, Y\left(\frac{1}{2} + \ell , \alpha r
\right) \, ,
\ee
where $J$ and $Y$ are Bessel functions of first and second kind,
respectively, and $A$ and $B$ are integration constants.

The solution of equation (\ref{Psi}) must be obtained by numerical
methods. We must thus introduce some values for the parameters $h$, $k$,
$\alpha^2$, and $m$. Following \cite{bwcoll2} we choose $h$ to be negative
so that our description of the system begins at a negative time and
evolves to a positive time. In order to simplify the calculation we choose
the simplest positive values for $k$ and $\alpha^2$, $k=|h|$ and
$\alpha^2=1$, as well as a positive scalar field mass $m$.

The potential term $(\alpha^2 +m^2 H^{1/2})$ is shown in
Fig.(\ref{pot}). We see that it is positive-definite although it possesses
some singular derivatives along the line $|y|=t$.

\begin{figure*}[htb!]
\begin{center}
\leavevmode
\begin{eqnarray}
\epsfxsize= 5.0truecm\rotatebox{-90}
{\epsfbox{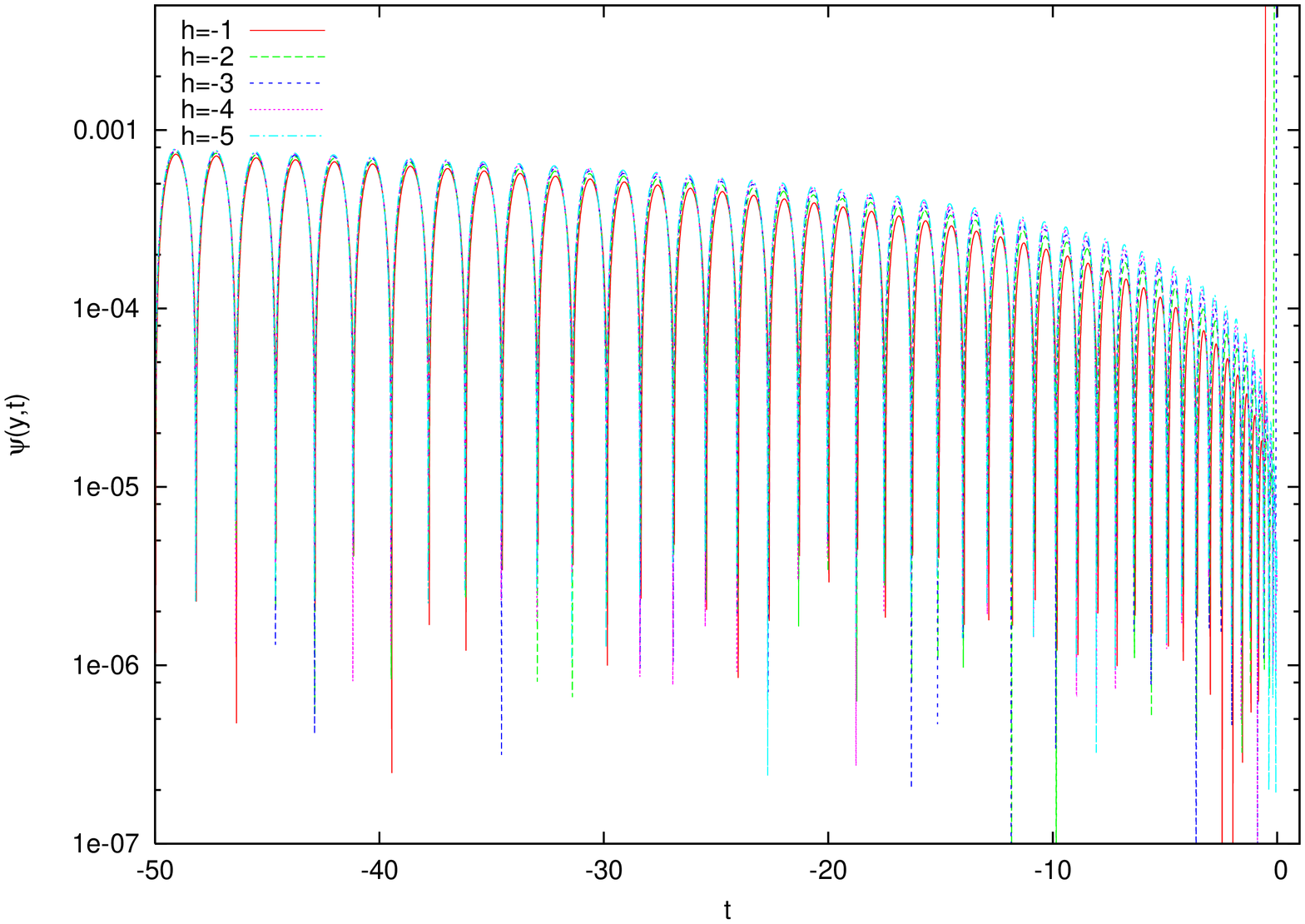}}\nonumber
\epsfxsize= 5.0truecm\rotatebox{-90}
{\epsfbox{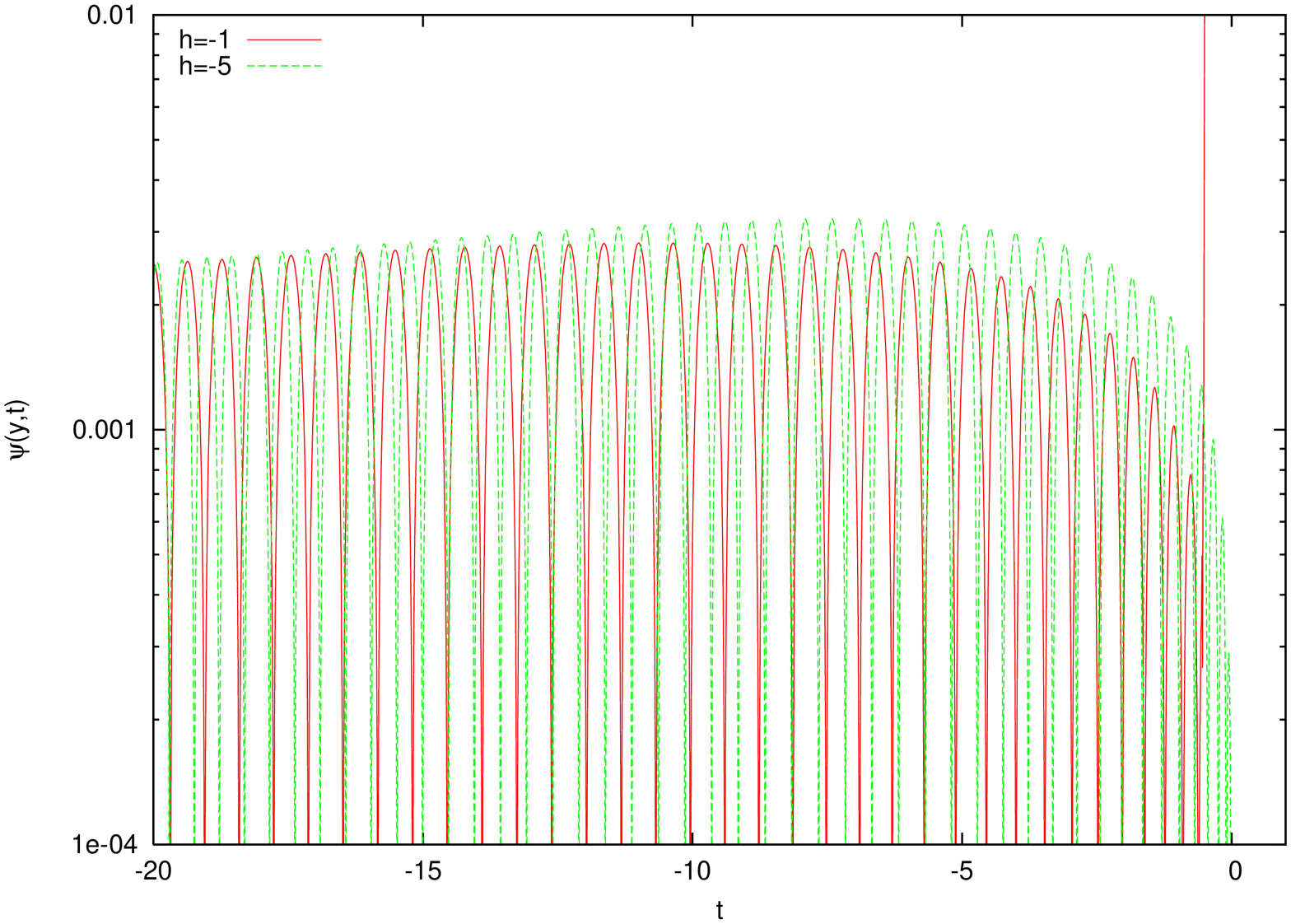}}\nonumber
\end{eqnarray}
\caption{Evolution of the massless (left) and massive (right) scalar field $\Psi$ at the position of the negative tension brane $y=0$ for several values of $h$.} 
\label{y0}
\end{center}
\end{figure*}

\begin{figure*}[htb!]
\begin{center}
\leavevmode
\begin{eqnarray}
\epsfxsize= 5.0truecm\rotatebox{-90}
{\epsfbox{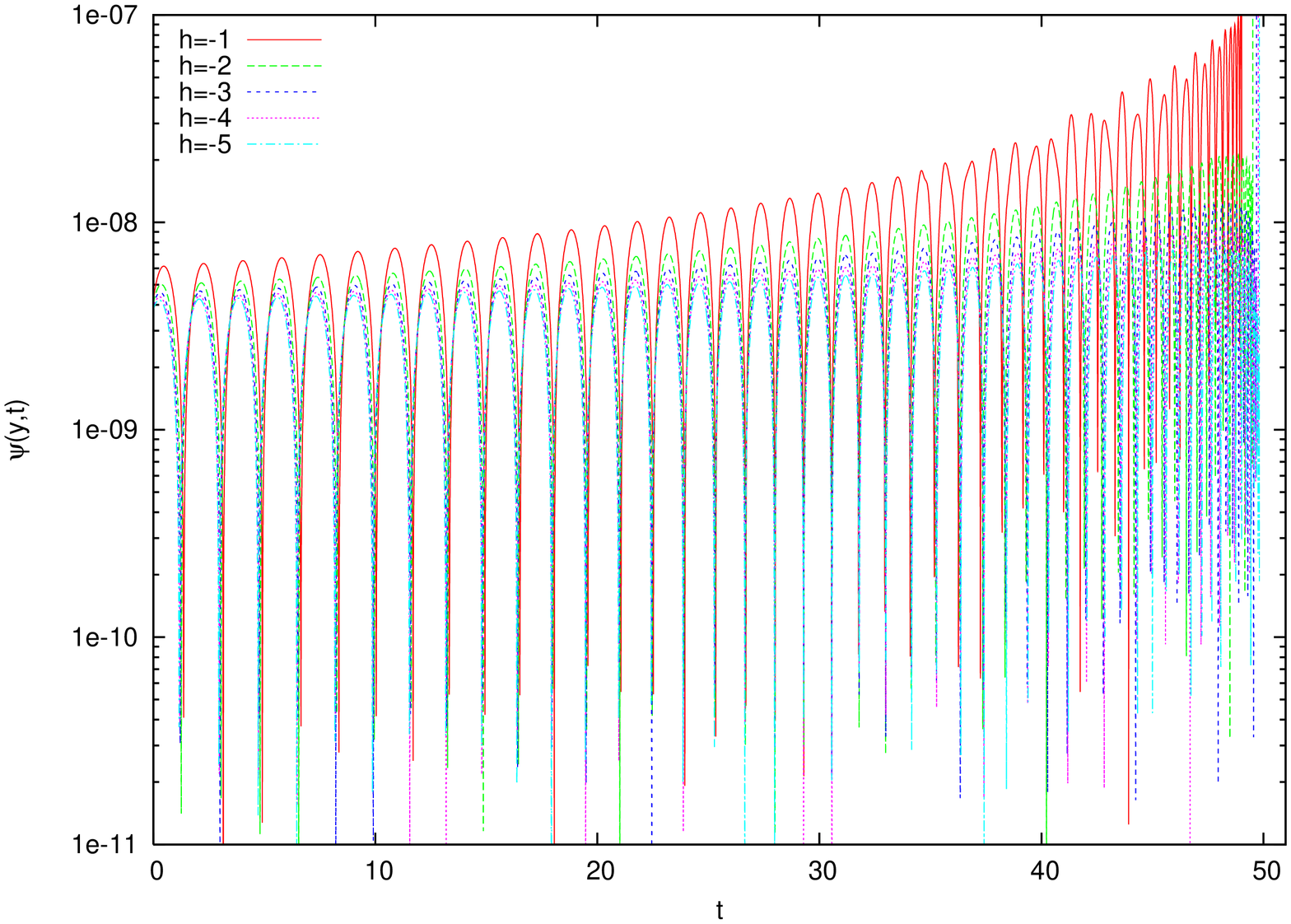}}\nonumber
\epsfxsize= 5.0truecm\rotatebox{-90}
{\epsfbox{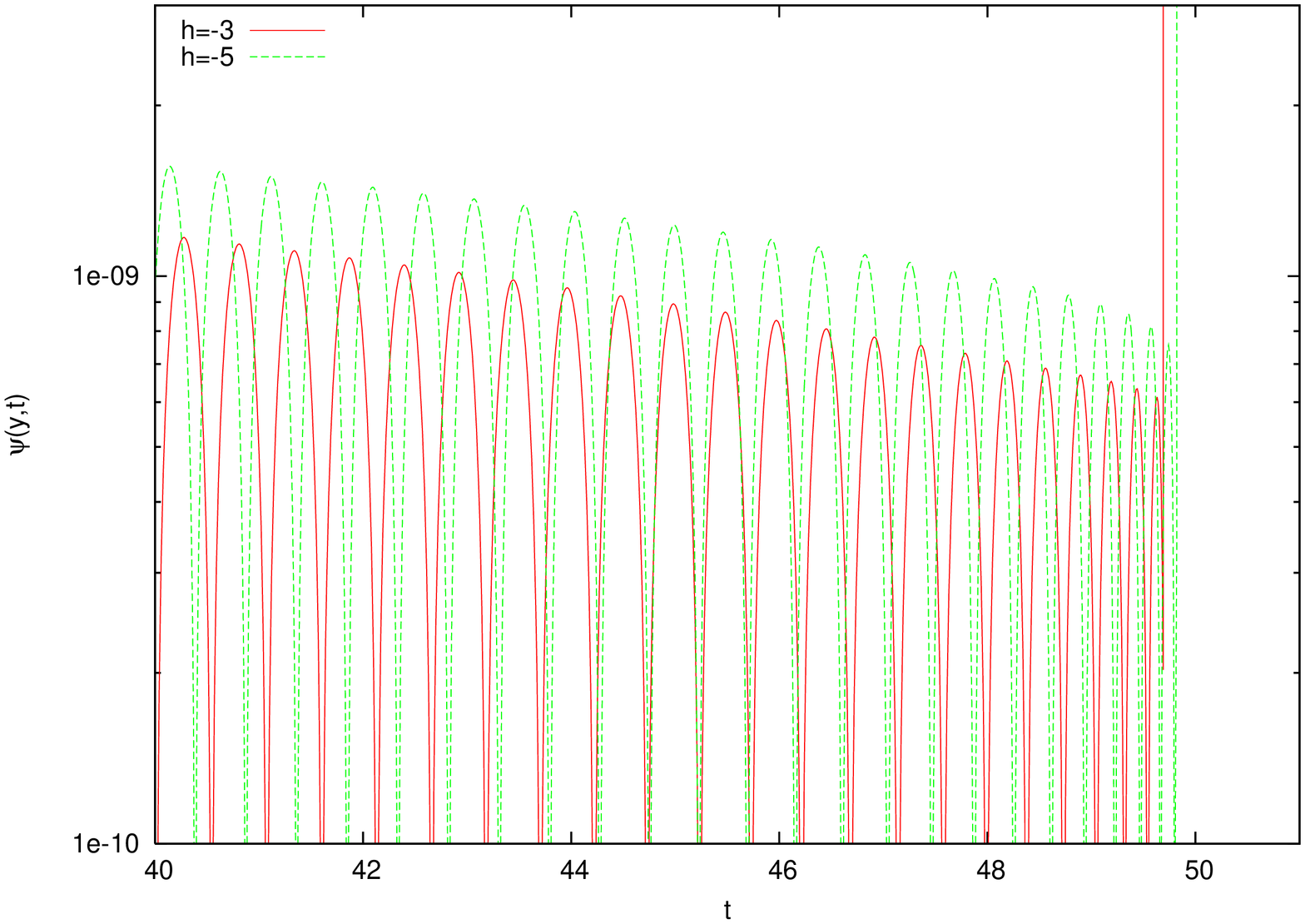}}\nonumber
\end{eqnarray}
\caption{Evolution of the massless (left) and massive (right) scalar field $\Psi$ at a bulk position $y=L/2$ between the two branes for several values of $h$.} 
\label{y50}
\end{center}
\end{figure*}

\begin{figure*}[htb!]
\leavevmode
\begin{eqnarray}
\epsfxsize= 5.2truecm\rotatebox{-90}
{\epsfbox{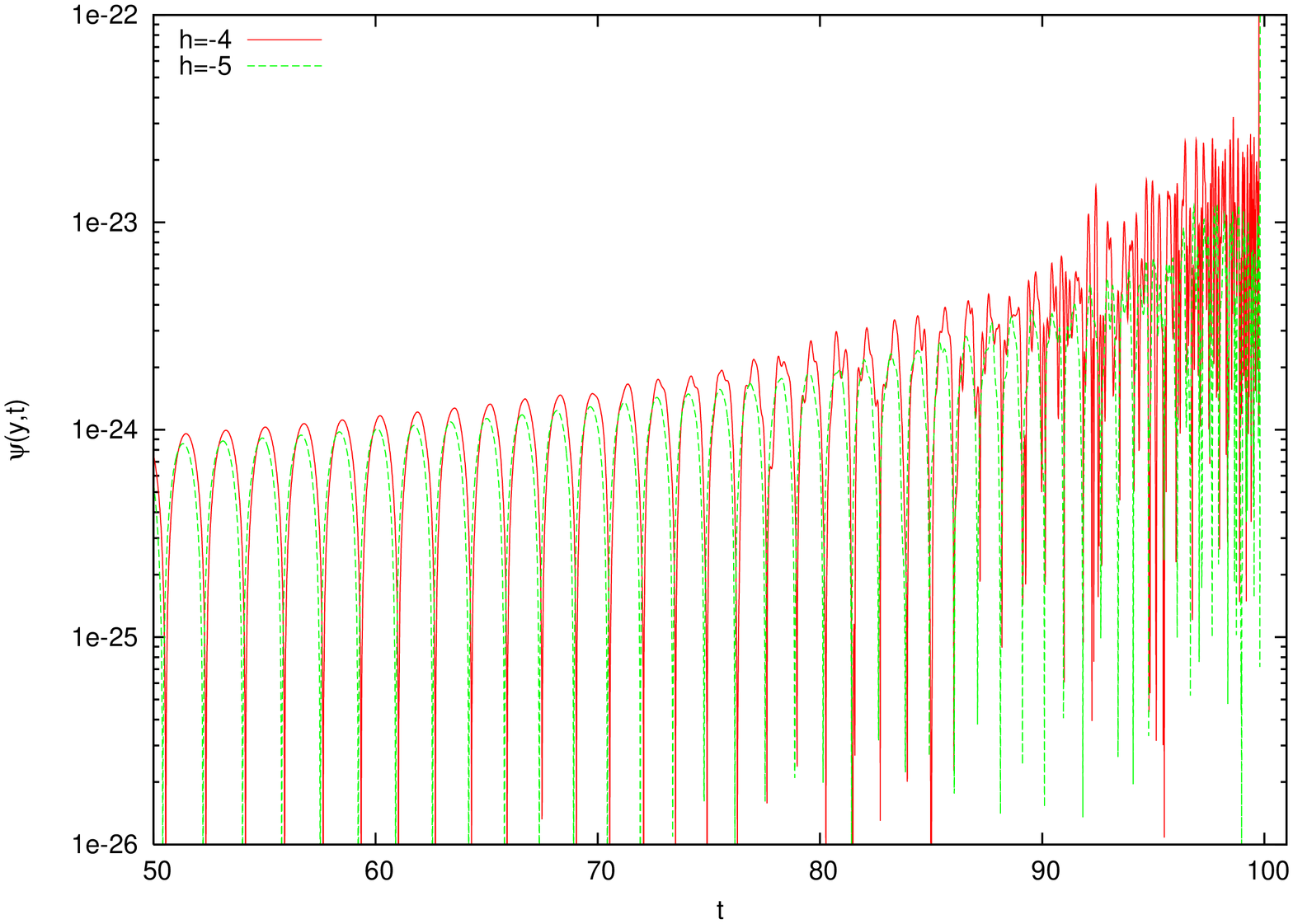}}\nonumber
\epsfxsize= 5.2truecm\rotatebox{-90}
{\epsfbox{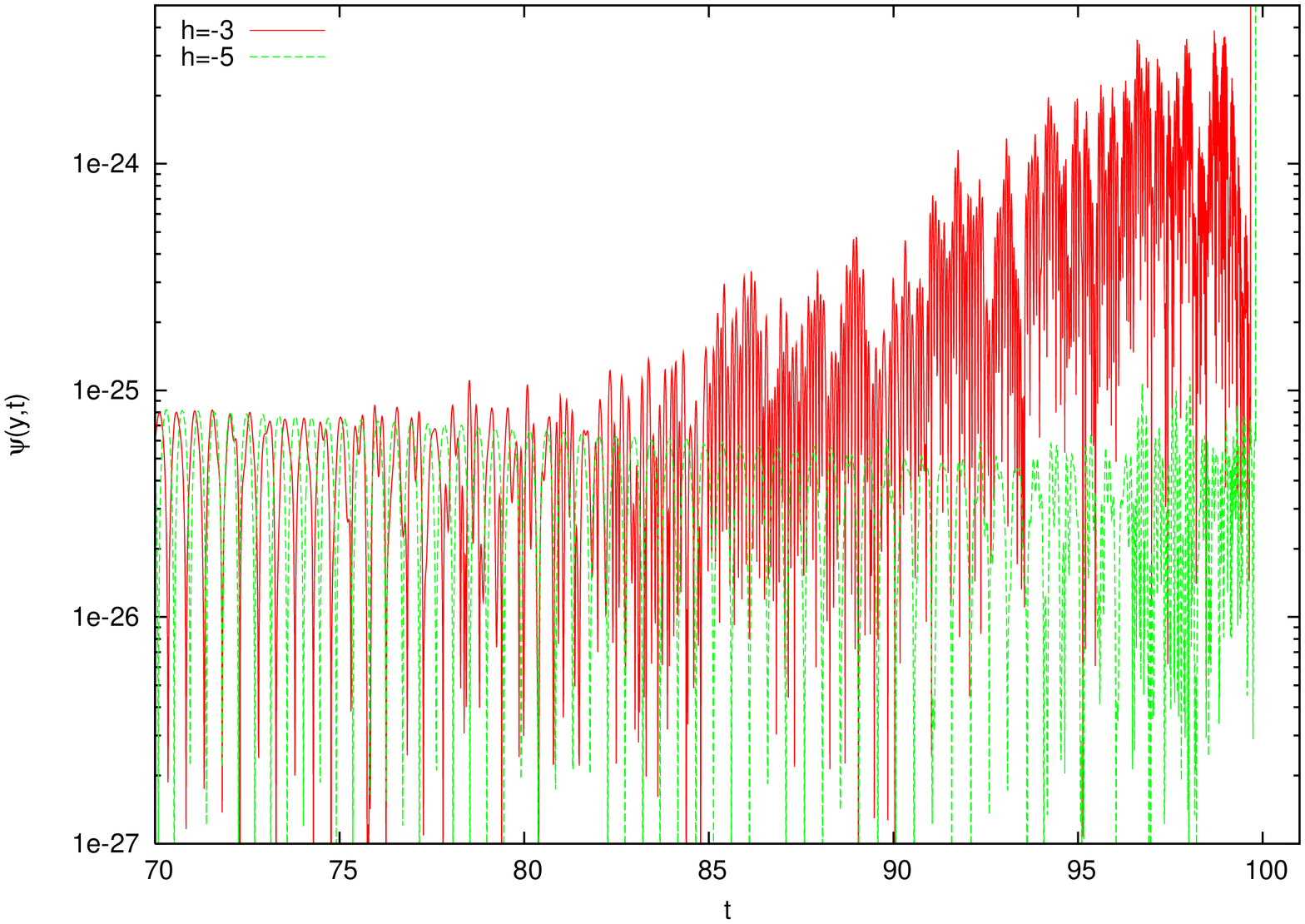}}\nonumber&&
\end{eqnarray}
\caption{Evolution of the massless (left) and massive (right) scalar field $\Psi$ at the position of the positive tension brane $y=L$ for several values of $h$.} 
\label{y100}
\end{figure*}

Our results for the evolution of the field $\Psi$ are shown in
Figs.\ref{y0}-\ref{y100}.

\begin{table}[t!]
\caption{Quasinormal Frequencies at $y=0$.} 
\begin{center}
\vspace{1cm}
\begin{tabular}{||c|c|c|c|c|c|c||}\hline\hline
${\bf |h|}$ & \multicolumn{2}{|c|}{{\bf $m=0$}} & \multicolumn{2}{|c|}
{{\bf $m=0.1$}} & \multicolumn{2}{|c||} {{\bf $m=2$}} \\ \hline \hline 
${\bf k}$ & ${\bf \omega_R}$&${\bf \omega_I}$&${\bf \omega_R}$&${\bf
  \omega_I}$&${\bf \omega_R}$&${\bf \omega_I}$ \\ \hline\hline 
1 & 1.428 & -0.0023 & 3.452 & 0.0015 & 6.411 & 0.0010 \\ \hline
2 & 1.428 & -0.0021 & 4.028 & 0.0016 & 7.662 & 0.0025 \\ \hline
3 & 1.428 & -0.0021 & 4.363 & 0.0014 & 8.491 & 0.0026 \\ \hline
4 & 1.428 & -0.0020 & 4.689 & 0.0017 & 8.976 & 0.0022 \\ \hline
5 & 1.428 & -0.0020 & 4.909 & 0.0016 & 9.520 & 0.0028 \\
\hline\hline
\end{tabular} 
\end{center}
\label{t0}
\end{table}

\begin{table}[t!]
\caption{Quasinormal Frequencies at $y=L/2=50$.}
\begin{center}
\begin{tabular}{||c|c|c|c|c|c|c||}\hline\hline
${\bf |h|}$ & \multicolumn{2}{|c|}{{\bf $m=0$}} & \multicolumn{2}{|c|}
{{\bf $m=0.1$}} & \multicolumn{2}{|c||} {{\bf $m=2$}} \\ \hline \hline 
${\bf k}$ & ${\bf \omega_R}$&${\bf \omega_I}$& ${\bf \omega_R}$& ${\bf
  \omega_I}$& ${\bf \omega_R}$& ${\bf \omega_I}$ \\ \hline\hline 
1 & 1.293 & -0.0005 & 3.740 & 0.0006 & 7.140 & 0.0013 \\ \hline
2 & 1.288 & -0.0005 & 4.363 & 0.0008 & 8.491 & 0.0010 \\ \hline
3 & 1.293 & -0.0006 & 4.760 & 0.0008 & 9.240 & 0.0017 \\ \hline
4 & 1.293 & -0.0006 & 5.150 & 0.0003 & 10.134 & -0.0025 \\ \hline
5 & 1.293 & -0.0006 & 5.417 & 0.0007 & 10.472 & 0.0022 \\
\hline\hline
\end{tabular} 
\end{center} 
\label{t50}
\end{table}

\begin{table}[t!]
\caption{Quasinormal Frequencies at $y=L=100$.} 
\begin{center}
\begin{tabular}{||c|c|c|c|c|c|c||}\hline\hline
${\bf |h|}$ & \multicolumn{2}{|c|}{{\bf $m=0$}} & \multicolumn{2}{|c|}
{{\bf $m=0.1$}} & \multicolumn{2}{|c||} {{\bf $m=2$}} \\ \hline \hline 
${\bf k}$ & ${\bf \omega_R}$&${\bf \omega_I}$& ${\bf \omega_R}$& ${\bf
  \omega_I}$& ${\bf \omega_R}$& ${\bf \omega_I}$ \\ \hline\hline 
1 & 1.199 & 0.0076 & 3.927 & 0.0773 & 7.854 & 0.0863 \\ \hline
2 & 1.200 & 0.0639 & 4.620 & 0.0642 & 9.240 & 0.0799 \\ \hline
3 & 1.204 & 0.0582 & 5.150 & 0.0590 & 10.472 & 0.0838 \\ \hline
4 & 1.213 & 0.0548 & 5.512 & 0.0551 & 10.833 & 0.0232 \\ \hline
5 & 1.213 & 0.0519 & 5.818 & 0.0524 & 11.220 & 0.0271 \\
\hline\hline
\end{tabular} 
\end{center}
\label{t100}
\end{table}

Fig.\ref{y0} provides information on the stability of the system before
$t=0$, when the singularity develops on the negative tension brane. We
can see that for earlier times the system is stable. The oscillations show
a quasinormal phase where we can calculate quasinormal frequencies (see
Table \ref{t0}), which show a strong dependence on the value of the
parameters $|h|$ and $k$ when $m\not=0$. We notice that the real part of
the frequency increases with $|h|$ and $k$, as well as with $m$. Since we
calculated the frequencies in the stability range where the oscillations
are almost uniform, the imaginary part appears to be very small, and its
sign just suggests a slight change in the amplitude of the oscillation
(positive sign indicates an increase, while negative sign indicates a
decrease). However, as soon as the scalar field detects the formation of
the singularity, the oscillations increase in amplitude and frequency,
revealing the instability of the system.

Once the singularity has formed on the negative tension brane, it
travels along the extra dimension in the direction of the positive tension
brane. Analizing our results at a point between the two branes as shown in
Fig.\ref{y50} we observe that the frequency of the scalar field oscillation increases gradually until encountering the singularity. Table \ref{t50} shows some quasinormal frequencies calculated by taking the most stable oscillations.

After a time $t=kL/(-h)$ the singularity reaches the positive tension
brane. In Fig.\ref{y100} we see how unstable the system becomes at this
point. The quasinormal frequencies shown in Table \ref{t100} provide 
evidence of this behaviour, since their positive imaginary part becomes
larger than in the previous cases. The singularity eventually wraps both
branes and the space between them, which constitutes the bulk, smearing
the actual brane collision.

\section{Conclusions}

In this paper we have considered the time-dependent Ho\u rava-Witten
solution found by Chen {\it et al.} \cite{chen}. We studied the stability
of the model by introducing a massive scalar perturbation.

We found that before any singularity appears, {\it i.e.}, for $t<0$, the
model evolves without any instability. For such a time-dependent solution
it was possible to calculate quasinormal frequencies at different
positions along the extra coordinate. We chose 3 representative points to
show our results, namely, at the position of each brane and between
the two branes. The quasinormal frequencies show a strong dependence on the
parameters of the model for a massive scalar field. The larger the
parameters $|h|$ and $k$ are, or alternatively the larger the mass of the
scalar field $m$ is, the larger the real part of the fundamental frequency
is. This dependence disappears when the mass of the scalar field
vanishes.

At $t=0$ a curvature singularity forms on the negative tension
brane. This singularity travels along the extra dimension and hits the
positive tension brane at $t=kL/(-h)$, before the two branes actually
collide.

From $t=0$ to $t=kL/(-h)$ the scalar field oscillations increase
frequency and amplitude showing the instability generated by the curvature
singularity that finally envelopes all the spacetime.

Despite knowing the instability of this model that leads to a big crunch, we
have shown that it is possible to detect it. The evolution of the scalar field
used for this purpose can provide some information about the parameters
of the model and the extension of the extra dimension. 

A complete picture of the instability should also include gravitational
perturbations. However, since both the metric and
the potential are time-dependent, such a procedure is beyond the aims of
the present paper.

\bigskip
{\bf Acknowledgements:} We wish to thank Elcio Abdalla, Bin Wang, and Chi Yong
Lin for enlightening discussions. This work has been supported by Funda\c
c\~ao de Amparo \`a Pesquisa do Estado de S\~ao Paulo {\bf (FAPESP)},
Brazil, and Fondo Nacional de Desarrollo Cient\'{i}fico y Tecnol\'ogico
{\bf (FONDECYT)}, Chile, under grant 3070009.


\begin{thebibliography}{s40}

\bibitem{hw} P.Ho\u rava, E.Witten, {\it Nucl. Phys.} {\bf B460}, 506 (1996).

\bibitem{lukas} A. Lukas, B.A. Ovrut, K.S. Stelle, and D. Waldram, {\it
  Phys. Rev.} {\bf D59}, 086001 (1999).

\bibitem{banks} T. Banks, M. Dine, {\it Nucl. Phys.} {\bf B479}, 173 (1996).

\bibitem{stab} J.L. Lehners, K.S. Stelle, and P. Smyth, {\it
    Class. Quant. Grav.} {\bf 22}, 2589 (2005). 

\bibitem{ekp} J. Khoury, B.A. Ovrut, P.J. Steinhardt, and N. Turok, {\it
    Phys. Rev.} {\bf D64}, 123522 (2001). 

\bibitem{cycle} P.J. Steinhardt and N. Turok, {\it Phys. Rev.} {\bf D65},
126003 (2002).

\bibitem{abdetal}
E. Abdalla, B. Cuadros-Melgar, S.S. Feng, B. Wang {\it
  Phys. Rev. } {\bf D65} (2002) 083512; E. Abdalla, A. Casali,
B. Cuadros-Melgar {\it  Nucl. Phys.}  {\bf B644}  (2002) 201-222. 

\bibitem{bwcoll1} G.W. Gibbons, H. Lu, and C.N. Pope, {\it
    Phys. Rev. Lett.} {\bf94}, 131602 (2005). 

\bibitem{chen} W. Chen, Z.-W. Chong, G.W. Gibbons, H. Lu, C.N. Pope, {\it
Nucl. Phys.} {\bf B732}, 118 (2006).

\bibitem{jeanluc} J.L. Lehners, P. McFadden, and N. Turok, {\it Phys. Rev.} {\bf D75}, 103510 (2007). J.L. Lehners and N. Turok, [arXiv:0708.0743].

\bibitem{x} K.D. Kokkotas and B.G. Schmidt {\it Living Rev. Rel.} {\bf 2}
  (1999) 2; B. Wang {\it Braz. J. Phys.} {\bf 35} (2005) 1029.

\bibitem{shao} C.G. Shao, B. Wang, E. Abdalla, and R.K. Su {\it
    Phys. Rev.} {\bf D71} (2005) 044003; E. Abdalla, C.B.M.H. Chirenti, and
  A. Saa {\it Phys. Rev.} {\bf D74} (2006) 084029 and {\it JHEP} {\bf
    0710} (2007) 086.

\bibitem{bwcoll2} D.R. Brill, G.T. Horowitz, D. Kastor, and J.H. Traschen,
  {\it Phys. Rev.} {\bf D49}, 840 (1994); G.W. Gibbons, H. Lu, C.N. Pope,
  {\it Phys. Rev. Lett.} {\bf 94}, 131602 (2005). 

\end{thebibliography}
\end{document}